# High harmonic spectroscopy reveals anisotropy of the Mott to Charge-Density-Wave phase transition in TiSe$_2$

Igor Tyulnev[1], Lin Zhang[1], Lenard Vamos[1], Julita Poborska[1], Utso Bhattacharya[2], Ravindra W. Chhajlany[5], Tobias Grass[3,4], Samuel Mañas-Valero[6], Eugenio Coronado[6], Maciej Lewenstein[1,3], Jens Biegert[1,7]*

[1]ICFO - Institut de Ciencies Fotoniques, The Barcelona Institute of Science and Technology, Castelldefels (Barcelona), Spain
[2]Institute for Theoretical Physics, ETH Zurich, Zurich, Switzerland
[3]DIPC - Donostia International Physics Center, Paseo Manuel de Lardizábal 4, San Sebastián, Spain
[4]Ikerbasque - Basque Foundation for Science, Maria Diaz de Haro 3, Bilbao, Spain
[5]ISQI - Institute of Spintronics and Quantum Information, Faculty of Physics and Astronomy, Adam Mickiewicz University, Poznań, Poland
[6]Instituto de Ciencia Molecular (ICMol), Universitat de València, Paterna, Spain
[7]ICREA, Pg. Lluís Companys 23, Barcelona, Spain
*Corresponding author : jens.biegert@icfo.eu

This work explores the use of polarization-resolved high harmonic generation (HHG) spectroscopy to investigate the quantum phases and transitions in the correlated charge density wave (CDW) phase of TiSe$_2$. Unlike previous studies focusing on crystallographic changes, the research examines the reordering that occurs within the CDW phase as the material is cooled from room temperature to 14 K. By linking ultrafast field-driven dynamics to the material's potential landscape, the study demonstrates how HHG is sensitive to quantum phase transitions. The findings reveal an anisotropic component below the CDW transition temperature, providing new insights into the nature of this phase. The investigation highlights the interplay between linear and nonlinear optical responses and their departure from simple perturbative dynamics, offering a fresh perspective on correlated quantum phases in condensed matter systems.

## Introduction

Highly correlated quantum phases arise from many-body interactions between charge carriers and the lattice, allowing quantum phenomena to appear at macroscopic scales[1]. In systems like 1T-TiSe$_2$, these interactions facilitate the formation of bosonic quasiparticles via phonon mediation, much like Cooper pair formation in superconductivity[2,3]. A related phase, the charge density wave (CDW)[4,5], follows similar principles, where the lattice vector Q connects high-symmetry points in the Brillouin zone, enabling exciton formation and a 2 x 2 x 2 commensurate periodic lattice distortion below ~200 K[6–10]. Recent studies[11–13] provide evidence for exciton condensation[14–16] in TiSe$_2$ when the exciton binding energy exceeds the bandgap, pointing to a novel phase of matter. Moreover, the material exhibits superconductivity upon Copper intercalation[17] or applied pressure[18], making it central to exploring high-temperature condensates. Anomalies such as chiral CDW stabilization[19,20], anisotropy[21], and strong CDW responses[6] add to the ongoing debate over the CDW mechanism— whether driven by electron-phonon interactions[9,22–24] or excitonic effects[11,25,26]. In this context, both electron-phonon (Jahn-Teller) and excitonic mechanisms have been proposed, and each aligns with experimental observations. Using high-harmonic generation spectroscopy, we investigate this prototypical phase transition from a new perspective, as it probes changes in correlations and symmetries[27].

## Discussion

### High Harmonic Spectroscopy

At the core of high harmonic generation (HHG) in solids[28–30] lies the precise response to charge currents through polarization and by crystal orientation. Analogous to the three-step model in gaseous HHG, charge carriers in solids are excited by strong optical fields, predominantly at band minima, and subsequently accelerated. The trajectories of these carriers, governed by field polarization and atomic potentials, allow recombination not only at the origin but at other lattice sites, enabling HHG spectroscopy to directly resolve crystallographic details[31,32]. The harmonic response, linked to optical properties and the complex dielectric function, is approximated through the Drude-Lorentz oscillator model, where resonances arise from interband transitions and their electron-hole populations.

Using the ponderomotive scaling for HHG cut-off energy, higher harmonics are driven at moderate intensities to preserve the correlated phase. To achieve this, we employ a mid-infrared optical parametric chirp pulse amplification (OPCPA) system[33,34], providing a stable optical field centered at 3.2 µm. This driving field impinges onto the sample in reflection geometry at a 45° incidence, and the reflected signal is spectrally analyzed. Harmonics are driven up to the 7th order with peak intensities reaching 40 GW/cm², while the linear response is captured through reflectivity measurements.

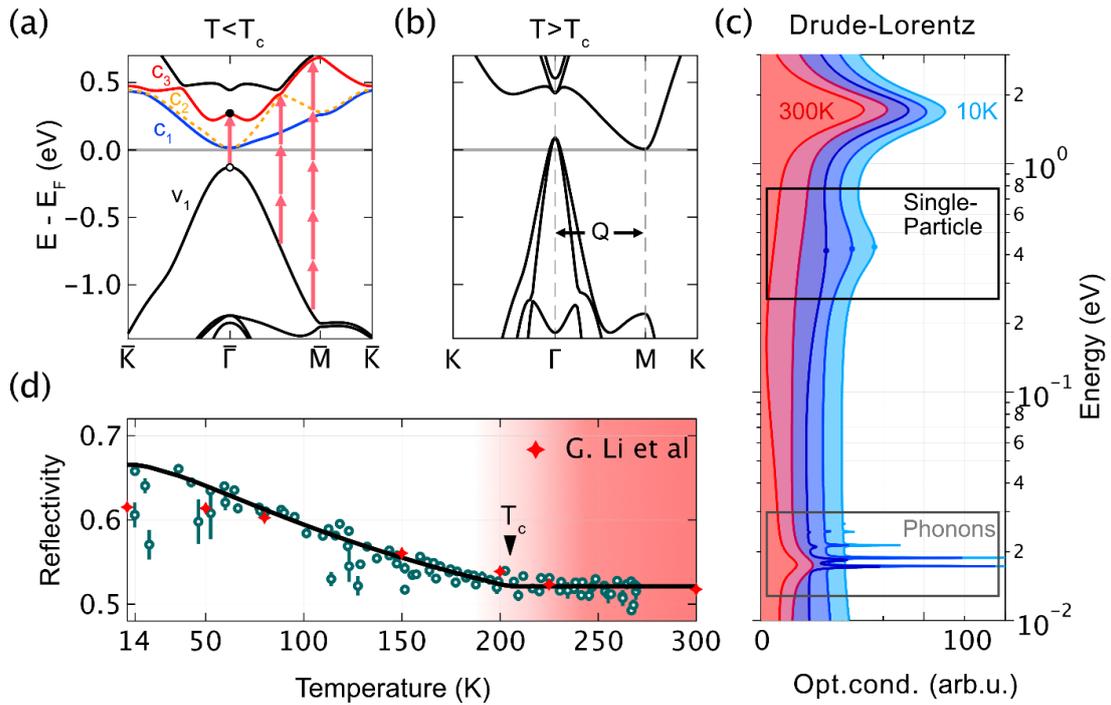

**Figure 1.** (a) and (b) Reduced and normal band structure of 1T-TiSe$_2$. The lattice vector Q, connecting the Γ and M points halfway across the Brillouin zone, leads to band back-folding and renormalization below $T_c$. The shifting of valence band $v_1$ and conduction band $c_3$ with temperature results in an IR transition of up to 0.4 eV in the CDW phase. Excited electron-hole pairs are driven by our mid-IR field, generating harmonics. (c) Optical conductivity calculated from a Drude-Lorentz fit to the measurements in Ref. 36. As temperature decreases, a resonance emerges below $T_c$, corresponding to the $v_1$ to $c_3$ transition. (d) Reflectivity of the fundamental laser frequency (0.39 eV) with a BCS fit and reflectivity values from Ref. 36.

Reflectivity data exhibits a phase-dependent response, remaining constant above the transition temperature ($T_c$) and deviating from linear scaling below $T_c$, resembling the mean-field form of the order parameter. A Bardeen-Cooper-Schrieffer-type fit[35] of the form of the form $\Delta^2 \propto \tanh^2\left(A\sqrt{T_c/T - 1}\right) + \Delta_0$, accurately captures this behavior (a = 0.82), yielding a transition temperature of $T_c$ = 205.0 ± 5.5 K, consistent with literature[6,7]. The resemblance to the CDW gap scaling becomes clear when examining the changes in optical properties near the photon energy of our fundamental field. As shown in Fig. 1c, the reflectivity is entirely governed by the temperature-dependent variations in optical conductivity or the dielectric function. In photoemission experiments, the CDW gap is indirectly inferred from shifts in the valence band ($v_1$) relative to the static "spectator" conduction bands ($c_1$ and $c_2$), which emerge during back folding below $T_c$; see Fig. 1a and b. This region is significantly affected by changes in the chemical potential and contributes to the low-frequency response, including the appearance of a plasma edge and phonon peaks; shown in Fig. 1c. Direct observation of the CDW gap between $v_1$ and $c_3$ is rare[36,37], as the empty $c_3$ band is often difficult to detect, though it offers a clear excitonic response. Probed at our fundamental frequency, this gap drives the emerging single-particle transition shown in Fig. 1c, measured optically through a Lorentzian resonance, with the band shift reaching up to 0.4 eV. The increased optical conductivity and reflectivity of TiSe$_2$, therefore, directly probe the band renormalization and CDW gap, as seen in Fig. 1d, showing up to a 22% rise in reflectivity at the lowest measured temperature of 14 K.

Beyond the single-particle response, we explore the harmonic yields as a function of temperature. Figure 2 reveals stark differences in the behavior of various harmonic orders, particularly when the driving field polarization aligns with the Γ-K direction. Approximating the response with the perturbative model for harmonic generation[38], changes in harmonic yield should directly reflect the temperature-dependent variations in nonlinear susceptibilities for each order. Since the dynamic laser parameters remain constant, the nonlinear susceptibility can be expressed as a product of linear susceptibilities at harmonic frequencies and the fundamental susceptibility raised to the respective power $\chi^{(a)} \propto \chi^{(1)}(a \cdot w_1)\left[\chi^{(1)}(w_1)\right]^a$.

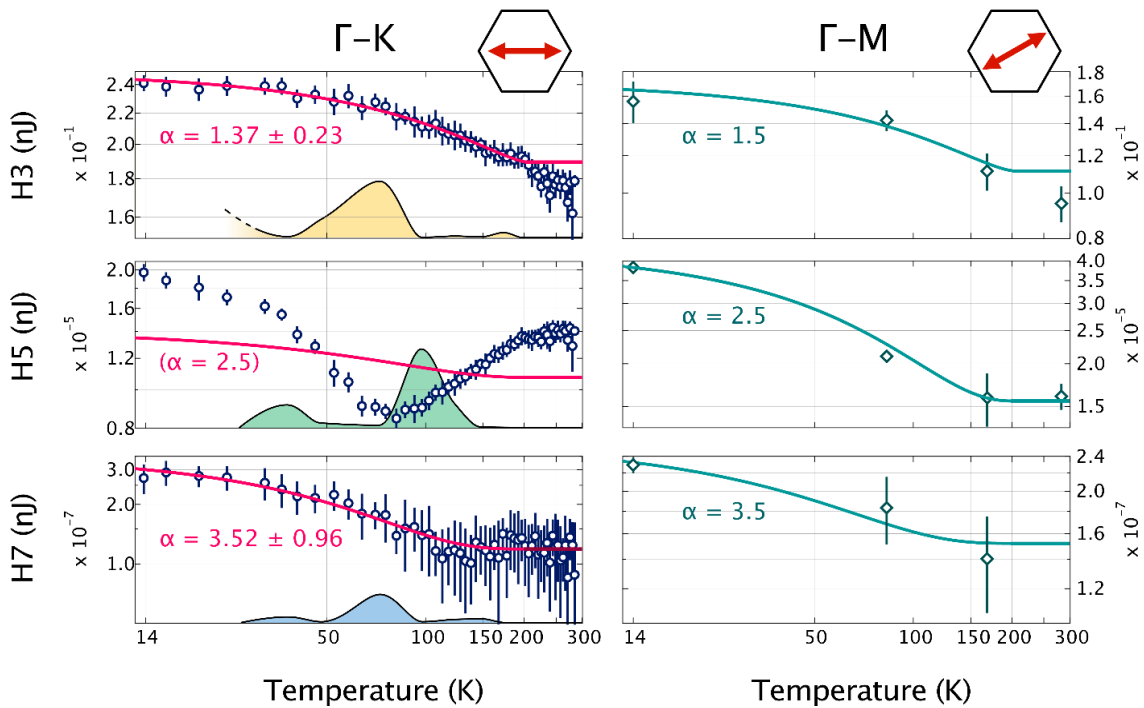

**Figure 2.** Measured harmonic intensity for the 3rd, 5th, and 7th orders as a function of temperature. The scans were conducted with the driving field polarization along the Γ-K direction, with square markers indicating additional measurements along the Γ-M direction, both plotted on a log-log scale. Shaded peaks

represent the optical conductivity calculated from the mean-field model at the harmonic frequencies: H3 (yellow), H5 (green), and H7 (blue). Solid lines denote the corresponding power-law scaling.

Power law fits were applied to each harmonic's temperature dependence. For H3, a fit of the form [$\propto (T_c - T)^\alpha$] yielded α = 1.37 ± 0.23 for T<$T_c$ consistent with the square-root scaling of the order parameter[25] raised to the third power. This scaling matches the perturbative description of the nonlinear susceptibility. For temperatures above $T_c$, deviations are attributed to the decreasing chemical potential, reflecting changes in electron-hole populations. The same fit for H7 reveals a square-root scaling to the seventh power with the temperature fit revealing α = 3.52 ± 0.96. In contrast, H5 exhibits unique sensitivity to the phase transition, with its yield decreasing below $T_c$ and reaching a minimum near 86 K. Comparing these findings with optical conductivity from our mean-field model (Fig. 2 shaded lines) at corresponding harmonic frequencies, we observe that significant changes in scaling for H5 occur near maxima in its optical properties. Further measurements, where HHG is driven along the Γ-M direction, reveal that the perturbative mechanism for H5 is restored, while other harmonics remain largely indifferent to the choice between Γ-M and Γ-K directions.

**High Harmonic Tomography of the CDW**

To further investigate the stark differences between harmonic orders and to understand the specific sensitivity of H5 to the CDW phase-transition, we conducted polarization scans across varying temperature regimes, controlling the input polarization with a half-wave plate. Figure 3 displays the behavior of H3, H5, and H7 as functions of the fundamental polarization angle. An angle of 90 degrees corresponds to the Γ-K direction, while 60 and 120 degrees align with the Γ-$M_1$ and Γ-$M_2$ directions, respectively. Near room temperature and above $T_c$, both H3 and H7 exhibit a peak for p-polarized input, while H5 splits into two symmetric peaks at 60-degree intervals, thus registering the hexagonal crystal symmetry.

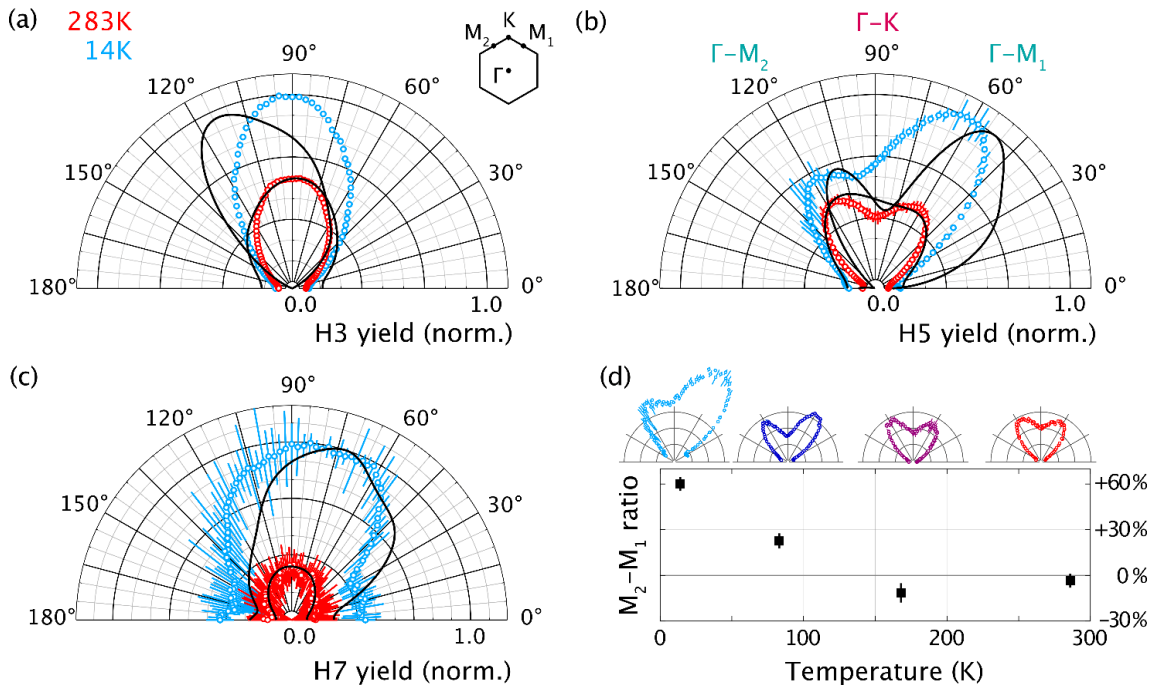

**Figure 3.** Harmonic yields for H3 (a), H5 (b), and H7 (c) as a function of the driving field polarization at 283 K and 14 K. The sketch illustrates the corresponding Brillouin zone orientation, with 90° indicating p-

polarization of the driving field along Γ-K. Black lines represent the simulated angular dependence of the harmonics based on the mean-field model, while dashed lines show the simulation adjusted for the angle offset relative to the experiment. (d) The relative ratio of H5 peaks at 60° and 120° as a function of temperature, with insets displaying the angular distribution.

As the sample is cooled below the phase transition, all harmonic peaks increase in intensity, with the contrast between H5's double peaks also becoming more pronounced, as seen in Fig. 3d, marked by the diminishing minimum between them. Below 86 K, H5 develops an asymmetry, with the peak at 60 degrees rising sharply at lower temperatures. This asymmetry also emerges in H7, where the center of mass shifts to the right, while H3 exhibits a slight but detectable rotation in the opposite direction. Comparing H5's temperature scaling with Fig. 2 for high-symmetry directions, the response along Γ-$M_1$ follows the mean-field behavior, but clear deviations occur along the Γ-K direction. Interestingly, the Γ-$M_2$ direction at 120 degrees, which should be symmetry-equivalent, remains nearly constant throughout the entire temperature range.

**Discussion**

These, at first glance unexpected, features reveal a set of observables that reflect the underlying microscopic changes that High Harmonic Spectroscopy sensitively reveals within the sample.
First, the overall increase in harmonic yield at lower temperatures highlights the phase transition, even within the nonlinear regime, demonstrating the coexistence of strong-field driving and the CDW phase. This points to the robustness of the CDW phase against moderate perturbations. Care was taken to limit the peak intensity and exposure to the sample, ensuring the experiment remained in the perturbative regime.
Notably, H5 exhibits a unique sensitivity to the crystal symmetry, differentiating between the zigzag (Γ-K) and armchair (Γ-M) directions. This can be attributed to the bandgap at the M point, which reaches up to 1.95 eV (Fig. 1a), aligning closely with the H5 energy of 1.96 eV, creating a near-resonant transition. In contrast, the K point's bandgap is slightly lower at 1.83 eV, favoring alignment along the Γ-M direction for H5. The increasing contrast below $T_c$, as shown in Fig. 3d, with the deepening Γ-K minimum and the enhanced Γ-M maximum, follows the renormalization of bands away from the Γ point.
Furthermore, the asymmetry between the theoretically equivalent Γ-$M_1$ and Γ-$M_2$ directions suggests an inequivalence between the lattice vectors $Q_1$, $Q_2$, and possibly $Q_3$, which drive the CDW formation. A closer look at Fig. 3d reveals two distinct asymmetry features in the H5 double peak. The first, a subtle negative asymmetry (where $M_2$ exceeds $M_1$), is already present at room temperature and becomes more pronounced around 150 K. Below 100 K, a strong positive asymmetry ($M_1 > M_2$) dominates, even influencing the Γ-K direction.
Turning to our mean-field model, which calculates the harmonic response for different polarization angles (see Methods), we observe excellent agreement in the high-temperature phase for all harmonics. The notable splitting of H5 along the crystal axes is accurately reproduced. To explore the origins of the asymmetry, we considered potential causes such as a slight rotational misalignment of the crystal during mounting. XRD analysis confirmed a 7° ± 2° offset between the crystal axis and the p-polarization of the laser, which could indeed account for the negative asymmetry observed at 283 K, as simulated for a 5° offset.
At high temperatures, this High Harmonic Spectroscopy study can determine crystal orientation and also identify the rotation direction. However, the pronounced low-temperature asymmetry prompted further investigation. While studies have proposed the chiral nature of the TiSe$_2$ CDW[19,39] and gyrotropic response[20], simulations introducing slight ellipticity in the driving field showed negligible differences, ruling out chirality of the field as a primary cause.

Ultimately, only an asymmetry between the translation vectors Q could produce such a pronounced effect. When the simulation assumed $\Delta Q_1 \neq \Delta Q_2 \neq \Delta Q_3$, a stronger asymmetry emerged, consistent with the experimental observation of H5 at 14 K. This anisotropic CDW dominates over the small offset effect seen at higher temperatures. Notably, the center of mass shifts of H5 and H7 to the right and H3 to the left are also reproduced, though slightly overestimated, supporting the idea of an inequivalence between the lattice vectors in the TiSe$_2$ CDW phase below 80-100 K.

A possible onset of strain at these temperatures could distort the CDW phase, potentially explaining our findings. While this is largely a lattice effect, it is worth noting that the Γ-K direction, which shows a significant signal increase at 14 K, aligns with the atomic displacement direction and the relevant phonon modes. Another explanation may lie in the backfolding of bands and their orbital character[40]. While Fig. 1a illustrates the three conduction bands $c_1$, $c_2$, and $c_3$ along the Γ-M path, they split along the M directions, and only Γ-M$_3$ has a band nearly resonant with H5. This strongly suggests that our experiments sensitively detects the concomitant hybridization between Se p-orbitals and Ti $d_{xy}$, $d_{yz}$, or $d_{zx}$ orbitals.

**Conclusion**

This work advances our insight into the quantum phases and transitions within the correlated charge density wave (CDW) phase of TiSe$_2$ by leveraging polarization-resolved high harmonic generation (HHG) spectroscopy. Distinct from prior research, which primarily focused on crystallographic changes, this study delves into the reordering occurring within the CDW phase as the material is cooled from room temperature to 14 K. By correlating ultrafast field-driven dynamics with the material's potential landscape, we demonstrate that HHG is exceptionally sensitive to quantum phase transitions and lattice dynamics. Our findings unveil an anisotropic component below the CDW transition temperature and reveal backfolding of the bands, shedding new light on the nature of this phase. This investigation underscores the complex interplay between linear and nonlinear optical responses and their deviation from simple perturbative dynamics, providing a novel perspective on the behavior of correlated quantum phases in condensed matter systems.

**Methods**

**Experiment**
The experiment was conducted using a home-built mid-infrared OPCPA laser system, delivering carrier-envelope phase (CEP) stable pulses of 100 fs duration at a repetition rate of 160 kHz and a central wavelength of 3.2 µm. By focusing the beam with a 150 mm lens, we achieved peak intensities of up to 40 GW/cm², allowing us to observe harmonic orders H3, H5, and H7 in a reflection geometry with a 45-degree incidence angle. The TiSe$_2$ sample, exfoliated from a larger bulk material to ensure a smooth surface, was placed in the vacuum chamber of a He cryostat. A UV-FS lens was used to image the beam into a spectrometer (OceanOptics Maya). To enhance the dynamic range for higher-order harmonics, a KG3 filter was employed to suppress the fundamental wavelength and attenuate H3. Alternatively, the reflected fundamental was measured using a power meter (Thorlabs S401C). The change of field amplitude when rotating the polarization was accounted for by the theory.

**Mean-field theory**
We employ a minimal phenomenological mean-field model to describe the charge density wave (CDW) order in TiSe$_2$. The Hamiltonian is expressed as:

$$H = \sum_{k}[\varepsilon_v(k) - \mu]d_{vk}^\dagger d_{vk} + [\varepsilon_c(k) - \mu]d_{ck}^\dagger d_{ck} + \sum_{Q,k} \Delta_{(-Q)} d_{vk+Q}^\dagger d_{ck} + \Delta_{(+Q)} d_{ck-Q}^\dagger d_{vk}.$$

Here, $\varepsilon_{v,c}(k)$ represent the valence and conduction band dispersions of TiSe$_2$, obtained from a tight-binding model[40]. The valence band, primarily composed of Se 4p orbitals, forms a hole pocket centered at the Γ point, while the conduction band, dominated by Ti 3d orbitals, creates three symmetry-equivalent electron pockets at the M points. The interaction leads to a triplet-Q CDW order, which couples the valence band hole pocket to the conduction band electron pockets, with wave vectors $|Q_1| = |Q_2| = |Q_3| = |\Gamma M|$.

The CDW order parameter[25] follows a mean-field temperature dependence, $\Delta(T) = \Delta_0\sqrt{1 - (T/T_c)^2}$, where $T_c = 200$ K. As a result of the CDW order, the Brillouin zone folds. In the reduced Brillouin zone (RBZ), the mean-field Hamiltonian can be recast as:

$$H = \sum_{k \in \text{RBZ}} \Psi_k^\dagger H(k) \Psi_k \text{ with } \Psi_k = \left(d_{vk}, d_{ck}, d_{vk+Q_1}, d_{vk+Q_1}, d_{vk+Q_2}, d_{vk+Q_2}, d_{vk+Q_3}, d_{vk+Q_3}\right)^T$$

**Optical Conductivity**
The optical conductivity of the material is expressed as:

$$\sigma_{ij}(\omega) = \frac{ie^2}{\omega V}\left(\sum_{k,mn} \frac{[f_m(k) - f_n(k)]H_{i,mn}(k)H_{j,nm}(k)}{i0^+ + \varepsilon_{km} - \varepsilon_{kn}} + \sum_{k,m}[U^\dagger(k)\,\partial_i\,\partial_j H(k)U(k)]_{mm} f_m(k)\right)$$
$$- \frac{ie^2}{V}\sum_{k,mn} \frac{[f_m(k) - f_n(k)]H_{i,mn}(k)H_{j,nm}(k)}{(i0^+ + \varepsilon_{km} - \varepsilon_{kn})(\omega + i0^+ + \varepsilon_{km} - \varepsilon_{kn})}$$

Where $H_{i,mn}(k) \equiv [U^\dagger(k)\partial_i H(k)U(k)]_{mn}$. $f_m(k) = 1/(1 + e^{\varepsilon_{km}/k_B T})$ is the Fermi-Dirac distribution with $\varepsilon_{km}$ being the eigenvalue of $H(k)$. We note that the conductivity has both the interband ($m \neq n$) and intraband ($m = n$) contribution, and the diagonal part of $[f_m(k) - f_n(k)]/(\varepsilon_{km} - \varepsilon_{kn})$ with $m = n$ must be interpreted as $\partial f(\varepsilon_{km})/\partial \varepsilon_{km}$.

**High Harmonic Generation**

We now calculate the high harmonic spectra of TiSe$_2$, which is a nonlinear optical process producing higher-order harmonics of the incident frequency when an intense laser field interacts with the material. We note that the coupling of a material to the laser field in dipole approximation amounts to the replacement $k \to k(t) = k + A(t)$ in the Bloch Hamiltonian using the Peierls substitution, where $A(t) = (A_x(t), A_y(t))$ with

$$A_x(t) = A(t)\sin\theta \sin(\pi/3 - \alpha) + \frac{\sqrt{2}}{2}A(t)\cos\theta \cos(\pi/3 - \alpha),$$

$$A_y(t) = A(t)\sin\theta \cos(\pi/3 - \alpha) - \frac{\sqrt{2}}{2}A(t)\cos\theta \sin(\pi/3 - \alpha),$$

and $A(t) = A_0 \sin^2\left(\frac{\omega t}{2n_{\text{cyc}}}\right)\sin(\omega t)$ is the vector potential of amplitude $A_0$, center frequency $\omega$, and cycle number $n_{\text{cyc}}$, projected onto the material plane (the laser field has an incident angle 45°). Here θ is the polarization angle and α is the crystal axis offset.

Using the velocity gauge equation of motion in the Bloch basis[41], the time-dependent Schrödinger equation can be written as

$$\frac{d}{dt}\rho(k,t) = -i[H(k + A(t); \{\Delta_{Q_i}(t)\}), \rho(k,t)]$$

for the density matrix $\rho(k,t)$ at momentum $k$, where $\Delta_{Q_i}(t) = (U/V)\sum_k \text{Tr}[\rho(k,t)d^\dagger_{vk+Q_i}d_{ck}]$ is the time-evolved CDW order parameter. Phenomenologically, we can choose different $U$ at different temperatures to obtain suitable strength of initial CDW order parameters. Here the initial state for each momentum is given by the Fermi-Dirac distribution, i.e., $\rho(k, 0) = 1/(e^{\beta H(k)} + 1)$. To include the phenomenological dephasing effect, after each time step $\delta t$ of the evolution governed by the above equation, we transform the density matrix into an adiabatic basis obtained by diagonalizing the instantaneous Hamiltonian $H(k + A(t); \{\Delta_{Q_i}(t)\})$, i.e., $\tilde{\rho}(k,t) = U^\dagger(k,t)\rho(k,t)U(k,t)$, apply the dephasing as $\tilde{\rho}_{mn}(k,t) \to \tilde{\rho}_{mn}(k,t)e^{-\delta t/\tau}$ for $m \neq n$ with τ being the dephasing rate, and then transform back to the Bloch basis.

We calculate the expectation value of the velocity operator in direction $j$

$$v_j(k,t) = \text{Tr}\left[\rho(k,t)\partial_{k_j}H(k + A(t); \{\Delta_{Q_i}(t)\})\right].$$

Denoting the polarization direction of the laser field as $n_A$ and $v_j(t) = (1/V)\sum_k v_j(k,t)$, then the high harmonic spectrum is given by the Fourier transform of velocity $P(\omega) = \omega^2|\text{FFT}[v \cdot n_A]|^2$.

**Acknowledgements**

J.B. acknowledges financial support from the European Research Council for ERC Advanced Grant "TRANSFORMER" (788218), ERC Proof of Concept Grant "miniX" (840010), FET-OPEN "PETACom" (829153), FET-OPEN "OPTOlogic" (899794), FET-OPEN "TwistedNano" (101046424), Laserlab-Europe (871124), MINECO for Plan Nacional PID2020–112664 GB-I00; QU-ATTO, 101168628; AGAUR for 2017 SGR 1639, MINECO for "Severo Ochoa" (CEX2019-000910-S), Fundació Cellex Barcelona, the CERCA Programme/Generalitat de Catalunya, and the Alexander von Humboldt Foundation for the Friedrich Wilhelm Bessel Prize. I.T. and J.B. acknowledge support from Marie Skłodowska-Curie ITN "smart-X" (860553). ICFO-QOT group acknowledges support from the European Research Council for AdG NOQIA; MCIN/AEI (PGC2018-0910.13039/501100011033, CEX2019-000910-S/10.13039 /

501100011033, Plan National FIDEUA PID2019-106901GB-I00, Plan National STAMEENA PID2022-139099NB, I00, project funded by MCIN/AEI/10.13039/501100011033 and by the "EU NextGenerationEU/PRTR" (PRTR-C17.I1), FPI); QUANTERA MAQS PCI2019-111828-2; QUANTERA DYNAMITE PCI2022-132919, QuantERA II Programme co-funded by EU Horizon 2020 program Grant No 101017733; Ministry for Digital Transformation and of Civil Service of the Spanish Government through the QUANTUM ENIA project call - Quantum Spain project, and by the European Union through the Recovery, Transformation and Resilience Plan - NextGenerationEU within the framework of the Digital Spain 2026 Agenda; Fundació Cellex; Fundació Mir-Puig; Generalitat de Catalunya (European Social Fund FEDER and CERCA program, AGAUR Grant No. 2021 SGR 01452, QuantumCAT\U16-011424, co-funded by ERDF Operational Program of Catalonia 2014-2020); Barcelona Supercomputing Center MareNostrum (FI-2023-3-0024); Funded by the EU. Views and opinions expressed are however those of the author(s) only and do not necessarily reflect those of the EU, European Commission, European Climate, Infrastructure and Environment Executive Agency (CINEA), or any other granting authority. Neither the EU nor any granting authority can be held responsible for them (HORIZON-CL4-2022-QUANTUM-02-SGA PASQuanS2.1, 101113690, EU Horizon 2020 FET-OPEN OPTOlogic, Grant No 899794, QU-ATTO, 101168628), EU Horizon Europe Program (This project has received funding from the EU's Horizon Europe research and innovation program under grant agreement No 101080086 NeQSTGrant Agreement 101080086 — NeQST); ICFO Internal "QuantumGaudi" project. U.B. is also grateful for the financial support of the IBM Quantum Researcher Program. T.G. acknowledges financial support from the Agencia Estatal de Investigación (AEI) through Proyectos de Generación de Conocimiento PID2022-142308NA-I00 (EXQUSMI), and that this work has been produced with the support of a 2023 Leonardo Grant for Researchers in Physics, BBVA Foundation. The BBVA Foundation is not responsible for the opinions, comments, and contents included in the project and/or the results derived therefrom, which are the total and absolute responsibility of the authors. RWC acknowledges support from the Polish National Science Centre (NCN) under the Maestro Grant No. DEC-2019/34/A/ST2/00081. We acknowledge the financial support from the European Union (ERC AdG Mol-2D 788222, FET OPEN SINFONIA 964396), the Spanish MCIN (2D-HETEROS PID2020-117152RB-100, co-financed by FEDER, and Excellence Unit 'María de Maeztu' CEX2019-000919-M) and the Generalitat Valenciana (PROMETEO Program, PO FEDER Program IDIFEDER/2018/061, a PhD fellowship to C.B.-C., and APOSTD-CIAPOS2021/215 to S.M.-V.). This study forms part of the Advanced Materials programme and was supported by MCIN with funding from European Union NextGenerationEU (PRTR-C17.I1) and by Generalitat Valenciana. S.M.-V. acknowledges the support from the European Commission for a Marie Sklodowska–Curie individual fellowship no. 101103355 - SPIN-2D-LIGHT. We thank Anna Palau and technical staff from The Institute of Materials Science of Barcelona (ICMAB-CSIC) for assistance with the XRD measurements.

**Author Statements**

J.B. conceived the project; I.T, L.V and J.P. conducted the measurements with support from J.B.; S.M.-V. and E.C. produced the samples; U.B., R.C., T.G. L.Z. and M.L. developed the mean-field theory and L.Z. conducted simulations; I.T., L.V., J.P. carried out the data analysis with support by J.B.; I.T. and J.B. wrote the manuscript with input from the authors.